\documentclass[12pt]{article}
\pdfoutput=1

\usepackage{color}
\usepackage{amssymb,amsmath,bm}
\usepackage{epsf}
\usepackage{epsfig}
\usepackage{afterpage}
\usepackage{longtable}
\usepackage{caption}
\usepackage{cite}
\usepackage{latexsym,mathrsfs}
\usepackage{graphics}
\usepackage{url}
\usepackage{paralist}
\usepackage{subcaption}
\usepackage{float}

\usepackage[dvipsnames]{xcolor}
\usepackage[linktoc=page,bookmarks=false,colorlinks=false,linkbordercolor=RoyalBlue,citebordercolor=ForestGreen,urlbordercolor=CornflowerBlue]{hyperref}

\usepackage[utf8x]{inputenc}

\usepackage{booktabs}

\newcommand{\IM}{{\rm Im}}
\newcommand{\RE}{{\rm Re}}
\newcommand{\GeV}{\, {\rm GeV}}
\newcommand{\tev}{\, {\rm TeV}}
\newcommand{\gev}{\, {\rm GeV}}
\newcommand{\mev}{\, {\rm MeV}}

\newcommand{\vcb}{|V_{cb}|}
\newcommand{\vtd}{|V_{td}|}
\newcommand{\vub}{|V_{ub}|}
\newcommand{\vts}{|V_{ts}|}

\newcommand{\kepe}{\kappa_{\varepsilon^\prime}}

\def\epe{\varepsilon'/\varepsilon}
\newcommand{\beq}{\begin{equation}}
\newcommand{\eeq}{\end{equation}}
\newcommand{\be}{\begin{equation}}
\newcommand{\ee}{\end{equation}}
\newcommand{\bi}{\begin{itemize}}
\newcommand{\ei}{\end{itemize}}
\newcommand{\ba}{\begin{array}}
\newcommand{\ea}{\end{array}}
\newcommand{\beqa}{\begin{eqnarray}}
\newcommand{\eeqa}{\end{eqnarray}}
\newcommand{\bea}{\begin{eqnarray}}
\newcommand{\eea}{\end{eqnarray}}
\newcommand{\beqn}{\begin{eqnarray}}
\newcommand{\eeqn}{\end{eqnarray}}

\newcommand{\eps}{\epsilon}

\definecolor{red}{cmyk}{0,1,1,0.4}

\def\kpn{K^+\rightarrow\pi^+\nu\bar\nu}
\def\klpn{K_{L}\rightarrow\pi^0\nu\bar\nu}

\newcommand{\wc}[3][{}]{\big[{\cal C}_{#2}^{#1}\big]_{#3}}

\newcommand{\op}[3][{}]{[{\cal O}_{#2}^{#1}]_{#3}}

\usepackage{fancyhdr}
\advance \headheight by 3.0truept       

\begin{document}

\begin{flushright}
    {AJB-23-2, \\
    LA-UR-23-20868}\\

\end{flushright}

\medskip

\begin{center}
{\Large\bf
  \boldmath{Kaon Physics Without New Physics in $ \varepsilon_K$}  }
\\[0.8 cm]
{\bf
  Jason~Aebischer$^{a}$,
    Andrzej~J.~Buras$^{b}$,
  Jacky Kumar$^{c}$
  }\\[0.5cm]
{\small
$^a$Physik-Institut, Universit\"at Z\"urich, CH-8057 Z\"urich, Switzerland \\[0.2cm]
$^b$TUM Institute for Advanced Study,
    Lichtenbergstr. 2a, D-85747 Garching, Germany \\[0.2cm]
$^c$Theoretical Division, MS B283, Los Alamos National Laboratory, Los Alamos, NM 87545, USA \\[0.2cm]
}
\end{center}

\vskip0.41cm

\begin{abstract}
  \noindent
  Despite the observation of significant suppressions of $b\to s\mu^+\mu^-$ branching ratios no clear sign of New Physics (NP) has been identified in $\Delta F=2$ observables $\Delta M_{d,s}$, $\varepsilon_K$ and the mixing induced CP asymmetries $S_{\psi K_S}$ and $S_{\psi\phi}$. Assuming negligible NP contributions to these observables allows to determine CKM parameters without being involved in the tensions between inclusive and exclusive determinations of $\vcb$ and $\vub$. {Furthermore this method avoids} the impact of NP on the determination of these parameters present likely in global fits. Simultaneously it provides SM predictions for numerous rare $K$ and $B$ branching ratios that are most accurate to date. Analyzing this scenario within $Z^\prime$ models we point out, following the 2009 observations of Monika Blanke and ours of 2020, that despite the absence of NP contributions to 
  $\varepsilon_K$, significant NP contributions to $\kpn$, $\klpn$, $K_S\to\mu^+\mu^-$, $K_L\to\pi^0\ell^+\ell^-$, $\epe$ and $\Delta M_K$
can be present. In the simplest scenario, this is
  guaranteed, as far as flavour changes are concerned, by a single non-vanishing  {\em imaginary} left-handed  $Z^\prime$ coupling  $g^L_{sd}$. This {scenario} implies very stringent correlations between {the} Kaon observables considered by us. In particular, the identification of NP in any of these
  observables implies automatically NP contributions to the remaining ones
  under the assumption of non-vanishing flavour conserving $Z^\prime$ couplings
  to $q\bar q$, $\nu\bar\nu$, and $\mu^+\mu^-$. A characteristic
  feature of this scenario is a strict correlation between   $\kpn$ and $\klpn$ branching ratios on a branch parallel to the Grossman-Nir bound.
  Moreover, $\Delta M_K$ is automatically suppressed as seems to be required by the results of the RBC-UKQCD lattice QCD collaboration. {Furthermore,} there is no NP contribution to $K_L\to\mu^+\mu^-$ which otherwise would bound NP effects in $\kpn$. Of particular interest are the correlations of $\kpn$ and $\klpn$ branching ratios and
  of $\Delta M_K$ 
  with the ratio $\epe$.  We investigate the impact of renormalization group
  effects in the context of the SMEFT
  on this simple scenario.
\end{abstract}

\thispagestyle{empty}
\newpage

\setcounter{tocdepth}{2}
\section{Introduction}
Despite a number of anomalies observed in $B$ decays,
it has recently been demonstrated \cite{Buras:2022wpw} that the
quark mixing observables
\smallskip
\be\label{loop}
{|\varepsilon_K|\,,\qquad \Delta M_s\,,\qquad \Delta M_d\,, \qquad S_{\psi K_S}}\,,\qquad 
S_{\psi \phi}\,,
\ee
\smallskip
\noindent
can be simultaneously described within the Standard Model (SM) without
any need for new physics (NP) contributions.
As these  observables contain by now only small hadronic uncertainties and are already  well measured this allowed to determine precisely the CKM matrix on
the basis of these observables alone without the need to face the tensions in
$\vcb$ and $\vub$ determinations from inclusive end exclusive tree-level decays
\cite{Bordone:2021oof,FlavourLatticeAveragingGroupFLAG:2021npn}.
Moreover, as pointed out in \cite{Buras:2022qip}, this also
avoids, under the assumption of negligible NP contributions to these observables, the impact of NP on the values of these parameters {, which are most likely} present in global fits. Simultaneously it provides SM predictions for numerous rare $K$ and $B$ branching ratios that are {the} most accurate to date. In this manner the size of the experimentally observed  deviations from SM predictions (the pulls) can be better estimated.

As over the {past} decades {the} flavour community expected significant impact of NP on
$\varepsilon_K$, $\Delta M_s$ and $\Delta M_d$, these findings, following dominantly from the 2+1+1 HPQCD lattice  calculations of $B_{s,d}-\bar B_{s,d}$ hadronic
matrix elements \cite{Dowdall:2019bea}, are not only surprising but also
putting very strong constraints on NP models attempting to explain the $B$ physics anomalies in question.

Concentrating on the $K$ system, which gained a lot attention recently \cite{Aebischer:2022vky,Goudzovski:2022scl,Aebischer:2022fld}, one could at first sight start worrying
that the absence of NP in a CP-violating observable like $\varepsilon_K$
would exclude all NP effects in rare decays governed by CP violation
 such as $\klpn$, $K_S\to\mu^+\mu^-$, $K_L\to\pi^0\ell^+\ell^-$ and also in the ratio $\epe$. Fortunately, these worries are premature. Indeed,
as pointed out already in 2009, in an important paper by Monika Blanke
\cite{Blanke:2009pq}, the absence of NP in $\varepsilon_K$ does not preclude
the absence of NP in these observables. This follows from the simple fact that
\be\label{EPSILONNEW}
(\varepsilon_K)_\text{BSM}\propto\left[(\RE(g_{sd})(\IM(g_{sd})\right],
\ee
where $g_{sd}$ is a  complex coupling present in a given NP model. Setting
$\RE(g_{sd})=0$, that is making this coupling {\em imaginary}, eliminates NP {contributions to} $\varepsilon_K$, while still allowing for sizable CP-violating effects in
rare decays and $\epe$. {This choice automatically eliminates the second solution considered in \cite{Blanke:2009pq} ($\IM(g_{sd})=0$), which is clearly less interesting.}

But there are additional  virtues of this simple NP scenario. It can
possibly explain the difference between the  SM
value for the $K^0-\bar K^0$ mass difference $\Delta M_K$ from  RBC-UKQCD \cite{Bai:2018mdv} and the data\footnote{A preliminary update including only statistical errors gives $(\Delta M_K)_\text{SM}=5.8(6) \times 10^{-15} \gev$
  \cite{Wang:2022lfq}.}
\be\label{NPDMK}
(\Delta M_K)_\text{SM}=7.7(2.1) \times 10^{-15} \gev,\qquad (\Delta M_K)_\text{exp} = 3.484(6) \times 10^{-15} \gev\,.
\ee
Indeed, as noted
already in \cite{Buras:2015jaq} and analyzed in the context of the SMEFT in
 \cite{Aebischer:2020mkv}, the suppression of $\Delta M_K$  is only possible in the presence
of new CP-violating couplings. This could appear surprising at first sight,
{since} $\Delta M_K$ is a CP-conserving quantity, but simply follows from the
fact that the BSM shift $(\Delta M_K)_\text{BSM}$ is proportional to the real part of the square of a complex
$g_{sd}$ coupling so that
\be\label{DeltaMK}
(\Delta M_K)_\text{BSM}= c~\RE[g_{sd}^2]= c\left[(\RE[ g_{sd}])^2-(\IM[ g_{sd}])^2\right], \qquad c>0\,.
\ee

With pure imaginary coupling, this suppression mechanism is very efficient.
The required negative contribution implies automatically NP contributions
to $\epe$ and also to rare decays  $K\to\pi\nu\bar\nu$,
$K_S\to\mu^+\mu^-$ and $K_L\to\pi^0\ell^+\ell^-$, provided this NP involves
non-vanishing flavour conserving $q\bar q$ couplings in the case of $\epe$
and non-vanishing $\nu\bar\nu$ and $\mu^+\mu^-$ couplings in the case
of {the} rare $K$ decays in question. In the case of $Z^\prime$ models
this works separately for left-handed and right-handed couplings and even
for the product of left-handed and right-handed couplings as long as all couplings are imaginary, {as discussed for instance in \cite{Aebischer:2020mkv,Aebischer:2019blw}}. This happens beyond the tree level and in any model
{in which} the flavour conserving couplings are real. Indeed, $\varepsilon_K$
  being a $\Delta F=2$ observable must eventually be proportional to $g^2_{sd}$. However, in order to see the implications of the absence of NP
  in $\varepsilon_K$ for Kaon physics, we concentrate in our  paper on
   $Z^\prime$ models.
  
But there is still one bonus in this scenario. With the vanishing
$\RE[g_{sd}]$ there is no NP contribution to $K_L\to\mu^+\mu^-$ which removes the constraint from this decay that can be important for $\kpn$.

In \cite{Aebischer:2020mkv} we have analyzed the observables listed in the abstract in NP scenarios with complex left-handed and right-handed $Z^\prime$ couplings to quarks in the context of the SMEFT. In view of the results of
\cite{Buras:2022wpw} it is of interest to repeat our analysis, {restricting the analysis to {an} imaginary $g_{sd}$  coupling as it eliminates NP contributions to $\varepsilon_K$ and also lowers the number of free parameters.}  Moreover, having already
CKM parameters determined in the latter paper, the correlations between various
observables are even more stringent than in  \cite{Aebischer:2020mkv} so that
this NP scenario is rather predictive.

However, it is also important to
 investigate the impact of renormalization group
 effects on this simple scenario in the context of the SMEFT to find out under which conditions  NP contributions to $\varepsilon_K$ are indeed negligible.

  Our paper is organized as follows. In Section~\ref{sec:2}, concentrating on left-handed couplings, we
  summarize our strategy that in contrast to our analysis in \cite{Aebischer:2020mkv} avoids the constraints from $K_{L}\to \mu^+\mu^-$
and $\varepsilon_K$. We refrain, with a few exceptions from listing the formulae for
observables entering our analysis as they can be found in
\cite{Buras:2015jaq,Aebischer:2020mkv}
and in {more general papers on $Z^\prime$ models in \cite{Buras:2012jb} and in
\cite{Buras:2012dp} that deals with 331 models.
In Section~\ref{sec:3}, we define the $Z^\prime$ setup, {and the observables analyzed by us} 
and briefly discuss the impact of SMEFT RG running on 
$\varepsilon_K$.
In Section~\ref{sec:4}
we present a detailed numerical analysis of
all observables listed above in the context of our simple scenario including
QCD and top Yukawa renormalization group effects. 
We conclude in Section~\ref{sec:5}.
\section{Strategy}\label{sec:2}
Our idea is best illustrated on the example of a new heavy $Z^\prime$ gauge boson
with {a} $\Delta S=1$ flavour-violating coupling $g_{sd}(Z')$ that is left-handed and purely imaginary\footnote{We suppress the index L for simplicity.}.
\be
   {\rm Re} g_{sd}(Z')=0\,, \qquad {\rm Im} g_{sd}(Z')\not=0\,.
   \ee
   As the tree-level contribution of this $Z^\prime$ is proportional to the imaginary part of the square of this coupling, it does not contribute at tree-level to
   $\varepsilon_K$ as stated in (\ref{EPSILONNEW}). However, it contributes to
   $\Delta M_K$ as seen in (\ref{DeltaMK}). It contributes also to several
   rare Kaon decays and also to $\epe${, for which the} NP contribution is just
   proportional to $ {\rm Im} g_{sd}(Z')$.

   Here we illustrate what happens on the basis of $\kpn$ and $\klpn$ decays.
   Their branching ratios are given as follows \cite{Buchalla:1998ba}
   \be \label{eq:BRSMKp}
  {\mathcal{B}(K^+\to \pi^+ \nu\bar\nu) = \kappa_+ \left [ \left ( \frac{{\rm Im} X_{\rm eff} }{\lambda^5}
  \right )^2 + \left ( \frac{{\rm Re} X_{\rm eff} }{\lambda^5}
  + \frac{{\rm Re}\lambda_c}{\lambda} P_c(X)  \right )^2 \right ] \, ,}
  \ee
  \be
  \label{eq:BRSMKL}
  {\mathcal{B}( K_L \to \pi^0 \nu\bar\nu) = \kappa_L \left ( \frac{{\rm Im}
    X_{\rm eff} }{\lambda^5} \right )^2 \, ,}
\ee
with $\kappa_{+,L}$  given by  \cite{Mescia:2007kn}
\begin{equation}\label{kapp}
\kappa_+={ (5.173\pm 0.025 )\cdot 10^{-11}\left[\frac{\lambda}{0.225}\right]^8}, \qquad \kappa_L=
(2.231\pm 0.013)\cdot 10^{-10}\left[\frac{\lambda}{0.225}\right]^8\,,
\end{equation}
and
\be\label{PCNNLO}
{X_{\rm SM}}=X(x_t)=1.462\pm 0.017\,, \qquad P_c(X)=(0.405\pm 0.024)\left[\frac{0.225}{\lambda}\right]^4\,.
\ee

In our model
\be\label{Xeff}
X_{\rm eff} = V_{ts}^* V_{td} X_{\rm SM} + X_{Z^\prime}\,, \qquad
 X_{Z^\prime}=\frac{g_{\nu\bar\nu}(Z')}{g^2_{\rm SM}M_{Z'}^2} g_{sd}(Z')\,,
\ee
where
\be\label{gsm}
g_{\text{SM}}^2=4\frac{G_F}{\sqrt 2}\frac{\alpha}{2\pi\sin^2\theta_W}=4 \frac{G_F^2 M_W^2}{2 \pi^2} =
1.78137\times 10^{-7} \gev^{-2}\,.
\ee

It should be noted that the SM one-loop function $X_{\rm SM}$ is real while
$X_{Z^\prime}$ is in our model purely imaginary. Thus the $Z^\prime$ contributes to $\kpn$ and $\klpn$ only through ${\rm Im} X_{\rm eff} $. The latter depends on  the sizes and
signs of the real $ g_{\nu\bar\nu}(Z')$ and $\IM  g_{sd}(Z^\prime)$ {couplings}.
Varying them the branching ratios for $\kpn$ and $\klpn$ are correlated on
the branch parallel to the Grossman-Nir bound, the so-called MB branch \cite{Blanke:2009pq}. They can either simultaneously increase or decrease relative to the SM predictions.
In the absence of NP in $\varepsilon_K$ the latter read  \cite{Buras:2021nns}
\be\label{BV}
{\mathcal{B}(\kpn)_\text{SM}= {(8.60\pm 0.42)}\times 10^{-11}\,,\quad
\mathcal{B}(\klpn)_\text{SM}={(2.94\pm 0.15)}\times 10^{-11}\,.}
\ee

Similarly the impact on $K_S\to\mu^+\mu^-$ and $K_L\to\pi^0\ell^+\ell^-$ is
only through the same $\IM  g_{sd}(Z^\prime)$  coupling, implying correlations with $\kpn$ and $\klpn$ branching ratios and also with $\epe$ subject to the values
of flavour conserving $Z^\prime$ couplings   to $q\bar q$, $\nu\bar\nu$ and $\mu^+\mu^-$.

  Consistent with our assumption of  negligible NP contributions in $\varepsilon_K$ and to the remaining $\Delta F=2$ observables in (\ref{loop}), we set
  the values of the CKM parameters to
\cite{Buras:2022wpw}
\be\label{CKMoutput}
{\vcb=42.6(4)\times 10^{-3}, \quad
\gamma=64.6(16)^\circ, \quad \beta=22.2(7)^\circ, \quad \vub=3.72(11)\times 10^{-3}\,,}
\ee
and consequently
\be\label{CKMoutput2}
{\vts=41.9(4)\times 10^{-3}, \qquad \vtd=8.66(14)\times 10^{-3}\,,\qquad
{\IM}\lambda_t=1.43(5)\times 10^{-4}\,,}
\ee
\be\label{CKMoutput3}
{\bar\varrho=0.164(12),\qquad \bar\eta=0.341(11)\,,}
\ee
where $\lambda_t=V_{ts}^*V_{td}$. The remaining parameters are given in Table~\ref{tab:input}. 
\begin{table}[!tb]
\center{\begin{tabular}{|l|l|}
\hline
$m_{B_s} = 5366.8(2)\mev$\hfill\cite{Zyla:2020zbs}	&  $m_{B_d}=5279.58(17)\mev$\hfill\cite{Zyla:2020zbs}\\
$\Delta M_s = 17.749(20) \,\text{ps}^{-1}$\hfill \cite{Zyla:2020zbs}	&  $\Delta M_d = 0.5065(19) \,\text{ps}^{-1}$\hfill \cite{Zyla:2020zbs}\\
{$\Delta M_K = 0.005292(9) \,\text{ps}^{-1}$}\hfill \cite{Zyla:2020zbs}	&  {$m_{K^0}=497.61(1)\mev$}\hfill \cite{Zyla:2020zbs}\\
$S_{\psi K_S}= 0.699(17)$\hfill\cite{Zyla:2020zbs}
		&  {$F_K=155.7(3)\mev$\hfill  \cite{Aoki:2019cca}}\\
	$|V_{us}|=0.2253(8)$\hfill\cite{Zyla:2020zbs} &
 $|\eps_K|= 2.228(11)\cdot 10^{-3}$\hfill\cite{Zyla:2020zbs}\\
$F_{B_s}$ = $230.3(1.3)\mev$ \hfill \cite{Aoki:2021kgd} & $F_{B_d}$ = $190.0(1.3)\mev$ \hfill \cite{Aoki:2021kgd}  \\
$F_{B_s} \sqrt{\hat B_s}=256.1(5.7) \mev$\hfill  \cite{Dowdall:2019bea}&
$F_{B_d} \sqrt{\hat B_d}=210.6(5.5) \mev$\hfill  \cite{Dowdall:2019bea}
\\
 $\hat B_s=1.232(53)$\hfill\cite{Dowdall:2019bea}        &
 $\hat B_d=1.222(61)$ \hfill\cite{Dowdall:2019bea}
\\
{$m_t(m_t)=162.83(67)\GeV$\hfill\cite{Brod:2021hsj} }  & {$m_c(m_c)=1.279(13)\GeV$} \\
{$S_{tt}(x_t)=2.303$} & {$S_{ut}(x_c,x_t)=-1.983\times 10^{-3}$} \\
    $\eta_{tt}=0.55(2)$\hfill\cite{Brod:2019rzc} & $\eta_{ut}= 0.402(5)$\hfill\cite{Brod:2019rzc}\\
$\kappa_\varepsilon = 0.94(2)$\hfill \cite{Buras:2010pza}	&
$\eta_B=0.55(1)$\hfill\cite{Buras:1990fn,Urban:1997gw}\\
$\tau_{B_s}= 1.515(4)\,\text{ps}$\hfill\cite{Amhis:2016xyh} & $\tau_{B_d}= 1.519(4)\,\text{ps}$\hfill\cite{Amhis:2016xyh}
\\
\hline
\end{tabular}  }
\caption {\textit{Values of the experimental and theoretical
    quantities used as input parameters. For future
updates see FLAG  \cite{Aoki:2021kgd}, PDG \cite{Zyla:2020zbs}  and HFLAV  \cite{Aoki:2019cca,HFLAV:2022pwe}.
}}
\label{tab:input}
\end{table}

As the SM prediction for $\epe$ is rather uncertain \cite{Buras:2022cyc,Aebischer:2020jto},
 we will, as in \cite{Aebischer:2020mkv}, fully concentrate on BSM contributions.\footnote{{General master formulae for the BSM effects can be found for example in \cite{Aebischer:2018quc,Aebischer:2018csl,Aebischer:2018rrz,Aebischer:2021hws}.}} Therefore in order to identify the pattern of BSM contributions to flavour observables
implied by allowed BSM contributions to $\epe$ in a transparent manner, we will proceed in our scenario by defining the parameter $\kepe$ as follows
\cite{Buras:2015jaq}
\be\label{GENERAL}
\frac{\varepsilon'}{\varepsilon}=\left(\frac{\varepsilon'}{\varepsilon}\right)^{\rm SM}
+\left(\frac{\varepsilon'}{\varepsilon}\right)^{\rm BSM}\,, \qquad
\left(\frac{\varepsilon'}{\varepsilon}\right)^{\rm BSM}= \kepe\cdot 10^{-3}\,, \qquad   0.0 \le \kepe \le 1.2\,.
\ee

In the case of $\varepsilon_K$ we allow only for very small NP contributions that
could be generated by RG effects despite setting the real part of
$g_{sd}(Z^\prime)$ to zero at the NP scale that we take to be equal
to $ M_{Z^\prime}$. Explicitly
\be
(\varepsilon)^{\rm BSM}=\kappa_\varepsilon\cdot  10^{-3}\,, \qquad -0.025\le |\kappa_\varepsilon|\le 0.025\,,
\ee
which amounts to $1\%$ of the experimental value. 

The SM predictions for $\kpn$ and $\klpn$ are
given in (\ref{BV}). For the remaining decays one finds with the CKM
parameters in (\ref{CKMoutput}) \cite{Buras:2022qip}
\newline
\begin{align}
  &\mathcal{B}(K_S\to\mu^+\mu^-)^{SD}_{SM} = (1.85\pm 0.12)\times 10^{-13}\,,\\
  &\mathcal{B}(K_L\to\pi^0e^+e^-)_{SM} = 3.48^{+0.92}_{-0.80}(1.57^{+0.61}_{-0.49})\times 10^{-11}\,,\\
  &\mathcal{B}(K_L\to\pi^0\mu^+\mu^-)_{SM} = 1.39^{+0.27}_{-0.25}(0.95^{+0.21}_{-0.20})\times 10^{-11}\,,
\end{align}
where for the $K_L\to\pi^0\ell^+\ell^-$ decays the numbers in parenthesis denote the destructive interference case. These results, that correspond to the CKM input in \eqref{CKMoutput}, differ marginally from the ones based on
\cite{Bobeth:2016llm, Isidori:2003ts, DAmbrosio:2017klp, Mescia:2006jd} used in
our previous paper. Note that the full $K_S\to\mu^+\mu^-$ branching ratio estimated in the SM including long-distance contributions is $\mathcal{B}(K_S\to\mu^+\mu^-)_{SM} = (5.2\pm 1.5)\times 10^{-12}$.

{The experimental status of these decays is given by\cite{LHCb:2020ycd, AlaviHarati:2000hs, Aaij:2017tia}}:
\begin{align}\notag
  &\mathcal{B}(K_S\to\mu^+\mu^-)_{\rm LHCb} < {2.1 \times 10^{-10}}\,,\quad \mathcal{B}(K_L\to\pi^0e^+e^-)_{exp} < 28 \times 10^{-11}\,,\\
  &\mathcal{B}(K_L\to\pi^0\mu^+\mu^-)_{exp} < 38 \times 10^{-11}\,.
\end{align}

\section{Setup} \label{sec:3}
{{\boldmath  \subsection{The $Z^\prime$ Model} } }
The interaction Lagrangian of a $Z'=(1,1)_0$ field and the SM fermions reads:
\begin{align}\label{eq:Zplag}
  \mathcal{L}_{Z'}=& -g_q^{ij} (\bar q^i \gamma^\mu q^j)Z'_{\mu}-g_u^{ij} 
(\bar u^i \gamma^\mu u^j)Z'_{\mu}-g_d^{ij} (\bar d^i \gamma^\mu d^j)Z'_{\mu}  \\\notag
  &-g_\ell^{ij} (\bar \ell^i \gamma^\mu \ell^j)Z'_{\mu}-g_e^{ij} (\bar e^i \gamma^\mu e^j)Z'_{\mu}\,.
\end{align}
\noindent
Here $q^i$ and $\ell^i$ denote left-handed $SU(2)_L$ doublets and $u^i$, $d^i$ and $e^i$ are right-handed singlets.
This $Z^\prime$ theory will then be matched at the scale $M_{Z^\prime}$ onto the SMEFT, generating 
 effective operators. The details of the matching onto SMEFT can be found in Ref.~\cite{Aebischer:2020mkv}.
As far as NP parameters are concerned we have the following real parameters
\be
M_{Z^\prime},\qquad  \IM(g_q^{21}),\qquad  g_\ell^{11}= g_\ell^{22} \qquad  g_u^{11}= -2 g_d^{11}.
\ee 
{The remaining parameters are set to zero.}
The latter flavour conserving couplings are required for the ratio $\epe$, for which significant differences
between various SM estimates can be found in the literature {\cite{Buras:2022cyc}}. The relation
between these couplings makes the electroweak penguin contributions to $\epe$ the dominant NP contributions. This is motivated by the
analyses in \cite{Buras:2015jaq,Aebischer:2020mkv} in which the superiority
of electroweak penguins over QCD penguins in enhancing $\epe$ has been
demonstrated. In fact the latter were ruled out by back-rotation effects \cite{Aebischer:2020lsx}.

{\boldmath
\subsection{The $\varepsilon_K$ due to RG Running}
}
\begin{figure}[htb]
\begin{center}
 \includegraphics[width=0.7\textwidth]{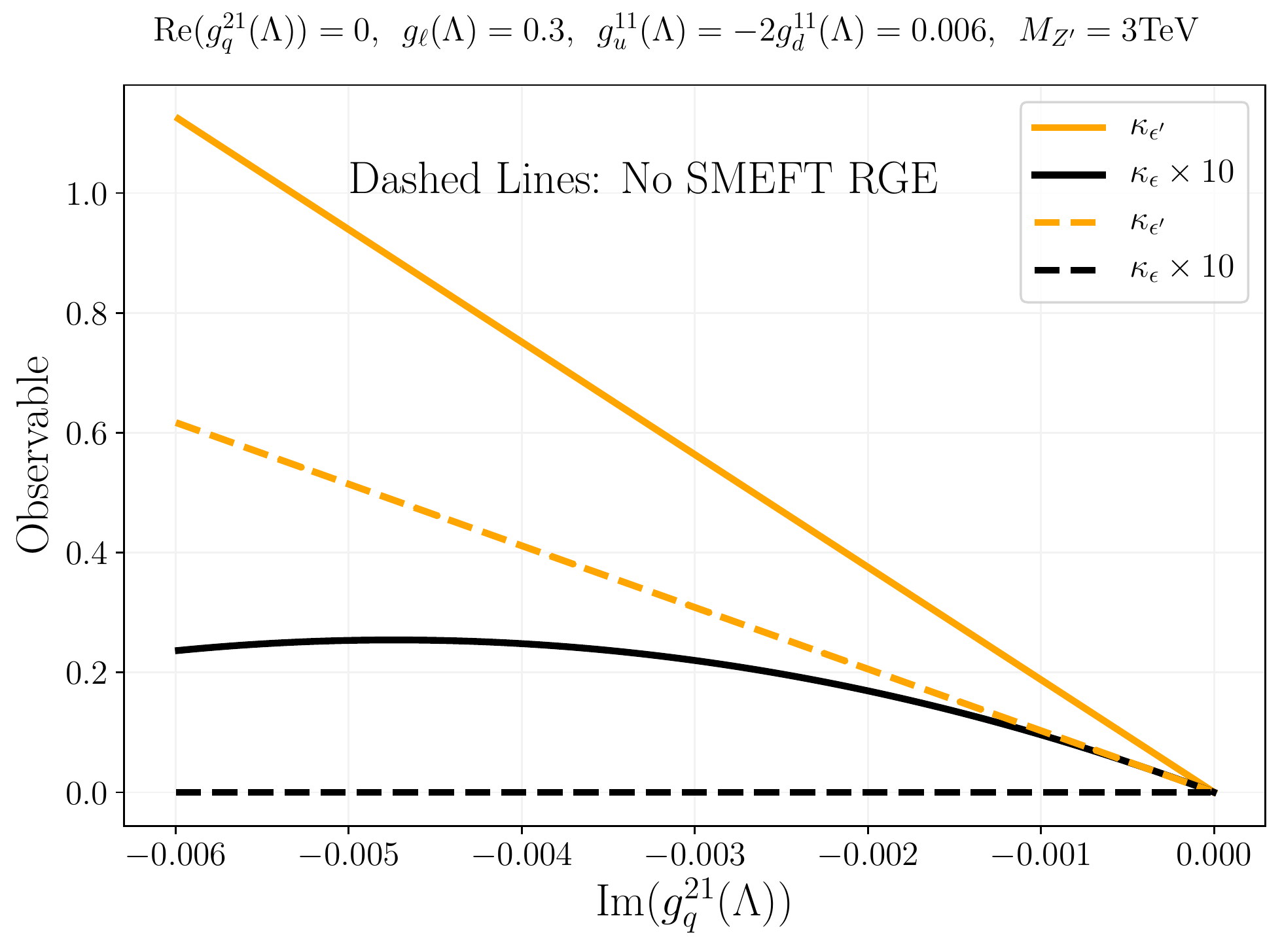}
\captionsetup{width=0.9\textwidth}
\caption{The  dependence of $\kappa_{\epsilon}$  and $\kappa_{\epsilon^\prime}$  on
Im$(g_{q}^{21})$ are shown. The Re$(q_q^{21})$ at scale $\Lambda$ is fixed to be zero.
The diagonal quark and lepton couplings are fixed as discussed in the text. 
Further, the solid and dashed lines correspond to $Z^\prime$
with and without SMEFT RG running effects. The WET (QCD+QED) running is
included in all cases. }
\label{fig:Imgsd-eps}
\end{center}
\end{figure}

In our scenario, at the scale of the $Z'$ mass, the $\Delta F=2$ operators contributing to $\varepsilon_K$ are assumed to be 
suppressed. However, this assumption may not always hold at the 
low energy scales and the contributing operators might still be generated due 
to SMEFT RG running.  We should make sure that this is not the case.
At the high scale, we have the following four-quark operators 
\bea
\op[(1)]{qq}{2121} &=& (\bar q_2 \gamma_\mu q_1) (\bar q_2 \gamma^\mu q_1)\,,  \\
\op[(1)]{qd}{2111} &=& (\bar q_2 \gamma_\mu q_1) (\bar d_1 \gamma^\mu d_1)\,.
\eea

In our scenario, at the scale $\Lambda=M_{Z'}$, the Wilson coefficient of the first operator is real and the second one is imaginary but it does not have the right flavour indices, so naively these operators
 should not affect $\varepsilon_K$. Through operator-mixing \cite{Jenkins:2013wua, Alonso:2013hga}, at the EW scale, {in the  leading-log approximation},
 we have 
 \bea 
 \Delta \wc[(1)]{qq}{2121}  (M_Z)&\approx&   {-}\frac{y_t^2}{16\pi^2}   \left (\lambda_t^{22} \wc[(1)]{qq}{2121}(M_{Z'}) + \Lambda_t^{11}  \wc[(1)]{qq}{2121}(M_{Z'})  \right)  \log{\left ( \frac{\Lambda}{M_Z} \right )}        \,,\\
\Delta \wc[(1)]{qd}{2111} (M_Z)&\approx&   {-} \frac{y_t^2}{16\pi^2}   \left (\lambda_t^{22} \wc[(1)]{qd}{2111}(M_{Z'}) + \lambda_t^{11}  \wc[(1)]{qd}{2111}(M_{Z'})  \right) \log{\left ( \frac{\Lambda}{M_Z} \right )}\,,  
 \eea
here, $\lambda_t^{11} \approx 8.3 \times 10^{-5}, \lambda_t^{22} \approx 1.7 \times 10^{-3}$. From the above two equations, it is clear that the RG running cannot induce the imaginary parts of $ \wc[(1)]{qq}{2121}$ and $ \wc[(1)]{qd}{2121} $ at $M_Z$. 

Therefore, our initial assumptions are stable under the SMEFT RGEs. However, we still find a small effect in 
$\varepsilon_K$ due to the back-rotation effect \cite{Aebischer:2020lsx}, which is basically due to the running of the SM down-type Yukawa couplings.
This is illustrated in Fig.~\ref{fig:Imgsd-eps}. {We observe that while the RG
  effects on $\varepsilon_K$ for this range of Im$(g_{q}^{21})$ are very small,
  they are large in the case of $\epe$.}

\subsection{Observables}
\bigskip

{In our numerical analysis we investigate the following quantities:}
\begin{eqnarray}\label{eq:kappas}
R_{\Delta M_K}    &=&  \frac{\Delta M_K^{BSM}} {\Delta M_K^{exp}} \,,\quad
R_{\nu\bar\nu}^+  =  \frac{\mathcal{B}(K^+ \to \pi^+ \nu \bar \nu)}{\mathcal{B}(K^+ \to \pi^+ \nu \bar \nu)_{SM}} \,,\quad
R_{\nu\bar\nu}^0 =  \frac{\mathcal{B}(K_L \to \pi^0 \nu \bar \nu)}{\mathcal{B}(K_L \to \pi^0 \nu \bar \nu)_{SM}}\,, \\\notag
R_{\mu^+\mu^-}^S  &=&  \frac{\mathcal{B}(K_S \to \mu^+\mu^-)_{\rm SD}}{\mathcal{B}(K_S \to \mu^+\mu^-)^{SD}_{SM}} \,,\quad R_{\pi\ell^+\ell^-}^0 =  \frac{\mathcal{B}(K_L \to \pi^0 \ell^+ \ell^-)}{\mathcal{B}(K_L \to \pi^0 \ell^+ \ell^-)_{SM}}\,,\qquad
\frac{\varepsilon'}{\varepsilon}\,.
\end{eqnarray}

Relative to \cite{Aebischer:2020mkv} the
constraint from $\mathcal{B}(K_L\to \mu^+ \mu^-)$ can be avoided because
there is no NP contribution to this decay in our scenario. In view of the recent
progress in \cite{Dery:2021mct} on the extraction of the short-distance contribution to the $K_S \to \mu^+\mu^-$ branching ratio, we will compare this time NP {contributions} to the
short-distance SM contribution and not to the full one that includes important long-distance {effects}.
{\boldmath
\section{Correlations between Kaon Observables: $|\kappa_{\varepsilon}| \le 0.025$}\label{sec:4}
}
\begin{figure}[t]
\begin{center}
 \includegraphics[width=1.\textwidth]{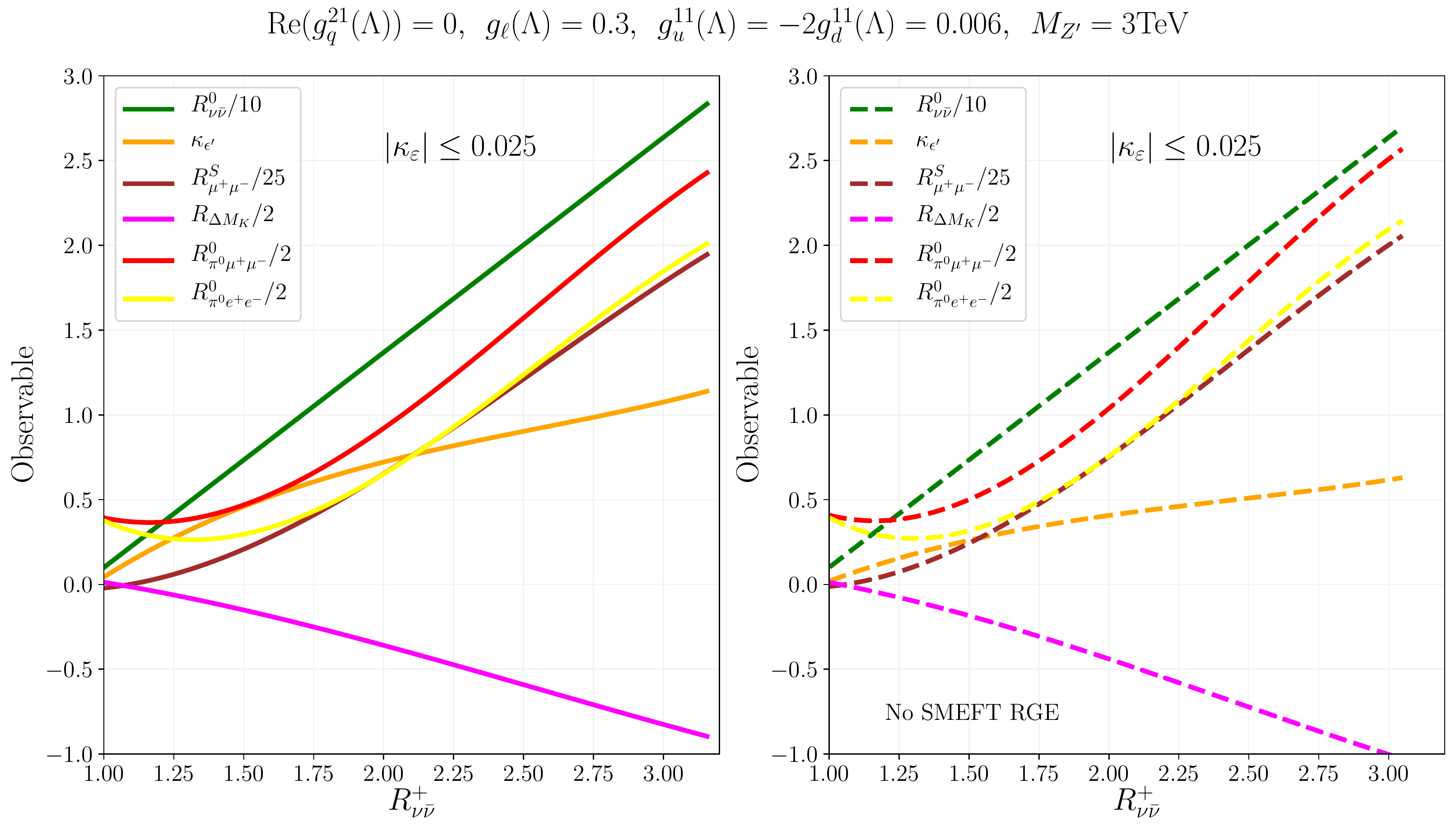}
\captionsetup{width=0.9\textwidth}
\caption{The correlations between the observable 
$R_{\nu \bar \nu}^+$ and various other Kaon observables are shown.
  The NP parameters {are given in \eqref{couplings3} and are the same as in
   } Fig.~\ref{fig:Imgsd-eps}.}
\label{Fig:2}
\end{center}
\end{figure}

In what follows we set the relevant $Z^\prime$ couplings at $\Lambda$ and its mass as in
\cite{Aebischer:2020mkv} to
\be\label{couplings3}
g_u^{11}=-2 g_d^{11}=6\times 10^{-3},\qquad g_\ell^{11}= g_\ell^{22}=0.3,
\qquad M_{Z^\prime}=3~\tev\,,
\ee
and
{\be\label{couplings10}
g_u^{11}=-2 g_d^{11}=6\times 10^{-2},\qquad g_\ell^{11}=g_\ell^{22}=3.0,
\qquad M_{Z^\prime}=10~\tev\,,
\ee
for simplicity we define $g_\ell^{11} = g_\ell^{22} = g_\ell $. While the lighter $Z^\prime$  is still in the reach of the LHC, the
 heavier one can only be discovered at a future collider.}

The relation between $g_u^{11}$ and $g_d^{11}$ assures that electroweak penguins
are responsible for the possible enhancement of $\epe$ with respect to the
SM value as expected within the Dual QCD approach \cite{Buras:2022cyc}.

For the numerical analysis the Python packages \texttt{flavio}\cite{Straub:2018kue}, \texttt{wilson}
\cite{Aebischer:2018bkb} and \texttt{WCxf} \cite{Aebischer:2017ugx} have been used,
in which the complete matching of the SMEFT onto the WET \cite{Aebischer:2015fzz,Jenkins:2017jig}, as well
as the full WET running \cite{Aebischer:2017gaw,Jenkins:2017dyc} are taken into account.
Note that some of the observables such as $R_{\Delta M_K}$, $R^{0}_{\pi \ell^+ \ell^-}$ are
not implemented in the public version of \texttt{flavio}. For these we have used our private codes.
In Fig.~\ref{fig:Imgsd-eps}  we show that for $M_{Z^\prime}=3~\tev$ with
\be\label{im}
-0.006\le \IM  (g_q^{21}) \le 0\,,
\ee
$\epe$ can indeed be significantly enhanced over its SM value while
keeping {the} NP impact on $\varepsilon_K$ below $1\%$ of the experimental value.
With the chosen quark couplings in (\ref{couplings3}) the negative values
of $\IM  (g_q^{21})$ are required to enhance $\epe$. For
  $M_{Z^\prime}=10~\tev$ the quark and lepton couplings have to be increased
to obtain similar {effects}. We observe significant RG effects in the case of $\epe$. 
Including them increases significantly the enhancement of $\epe$
  for a given  $\IM  (g_q^{21})$. On the other hand this effect is
very small in the case of $\varepsilon_K$.

An important test of this NP scenario will be the correlations between
  all observables discussed by us. This is illustrated in Figs.~\ref{Fig:2} and
\ref{Fig:3} for $M_{Z^\prime} =3~\tev$ and  $M_{Z^\prime} =10~\tev$, respectively.
We show there the dependence of various observables on the ratio
$R_{\nu\bar\nu}^+$ restricting  the values of
 $\IM (g_q^{21}) $ to the range in (\ref{im}) for which the NP effects in 
$\varepsilon_K$ are at most $1\%$. Of particular importance is the correlation between  $R_{\nu\bar\nu}^+$ and $R_{\nu\bar\nu}^0$.  As announced before the full action takes {part} exclusively on the MB branch parallel to the GN bound not shown in the plot. We observe {a} strong enhancement of $\klpn$ branching ratio.
Finding in the future the experimental values of both branching ratios
  outside the MB branch would indicate, among other possibilities, the presence
  of other particles which would affect ${\rm Re} X_{\rm eff}$.
While the remaining correlations are self explanatory, let us make the following
observations 

\begin{itemize}
\item
  The four ratios in Fig.~\ref{Fig:2} and \ref{Fig:3} that increase with increasing $R^+_{\nu \bar \nu}$ 
  are very strongly correlated with each other because  being CP-violating
  they depend only on $\IM (g_q^{21}) $. While $\epe$ and  $R_{\pi\mu^+\mu^-}^0$
  can be significantly
  enhanced, the enhancements of $R_{\nu\bar\nu}^0$ and $R_{\mu^+\mu^-}^S$ are huge
  making hopes that even a moderate enhancement of $R_{\nu\bar\nu}^+$ over the SM
  prediction will allow to observe $\klpn$ and $K_S\to\mu^+\mu^-$ in the coming
  years.
\item
The RG effects play a significant
role in $\kappa_{\epsilon^\prime}$ and $R_{\Delta M_K}$  but otherwise these effects are small. RG effects are simply larger in non-leptonic
decays than in semi-leptonic and leptonic ones.
\end{itemize}
The size of leptonic and semileptonic branching ratios depends on 
leptonic couplings but the correlations themselves do not depend
on them because for left-handed couplings the couplings to charged leptons and neutrinos must be the same due to the unbroken $\text{SU(2)}_L$ symmetry
in the SMEFT.

\begin{figure}[t]
\begin{center}
 \includegraphics[width=1.\textwidth]{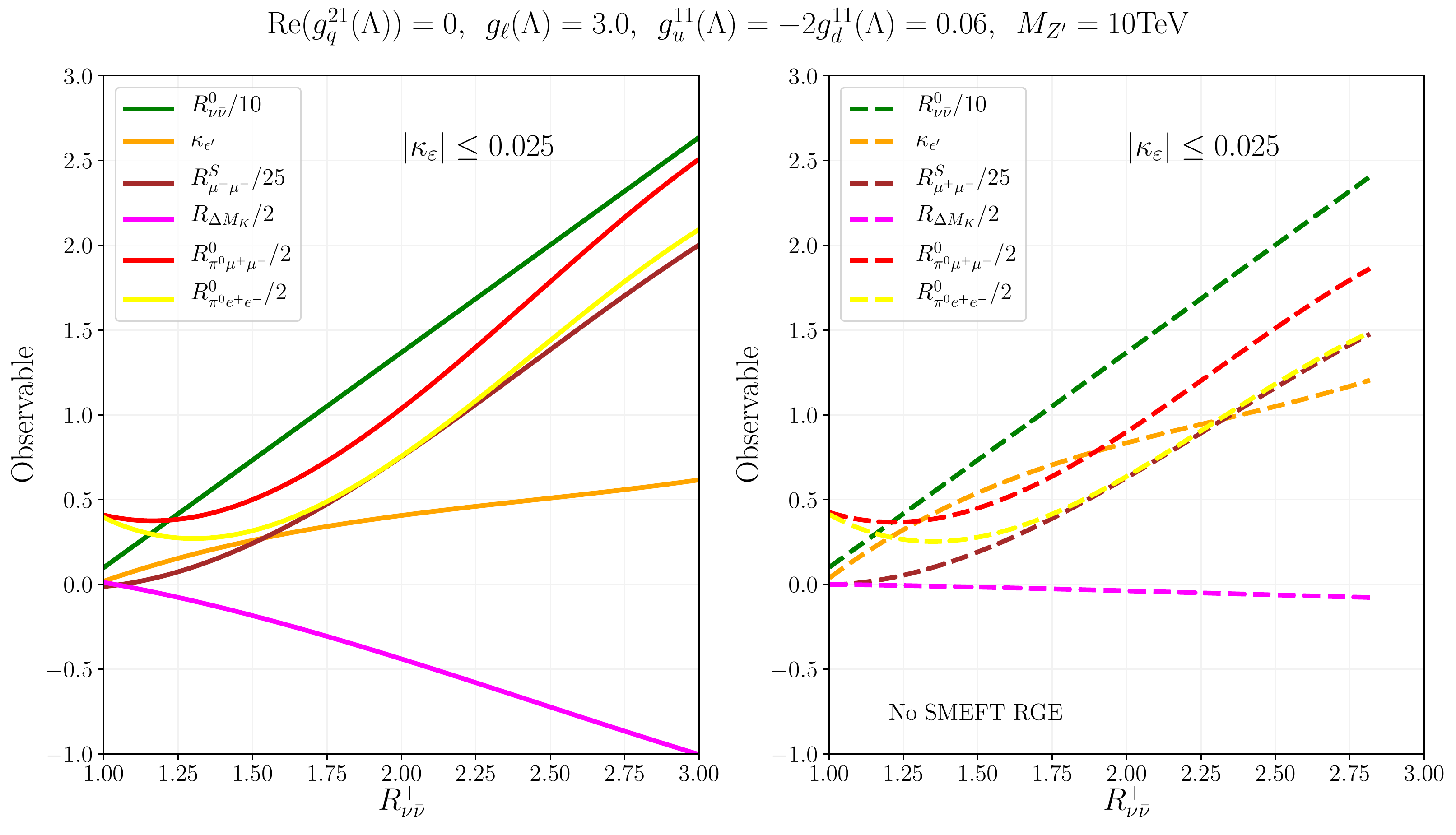}
\captionsetup{width=0.9\textwidth}
\caption{The correlations between the observable 
$R_{\nu \bar \nu}^+$ and various other Kaon observables are shown.
The NP parameters and other details are the same as given in \eqref{couplings10}.}
\label{Fig:3}
\end{center}
\end{figure}

In order to illustrate the power of correlations we anticipate the future
discovery of $Z^\prime$ with its mass $3\tev$ and the
measurement of NA62 collaboration resulting in
\be
R_{\nu\bar\nu}^+=1.50\, .
\ee
For $g_\ell=0.3$  we find then
\be
R_{\nu\bar\nu}^0 \approx 7.5, \qquad \kappa_{\epsilon^\prime} \approx 0.5,\qquad R_{\mu^+\mu^-}^S  \approx  5.0\,,
\ee
and
\be
R_{\Delta M_K} \approx  -0.25,\qquad R^0_{\pi^0 \mu^+ \mu^-} \approx 1.0 \qquad  R^0_{\pi^0 e^+ e^-} \approx 0.5.
\ee
Certainly, the results depend on the leptonic couplings. In the future they
could be determined from other processes, in particular from B decays.

  \section{Summary and Outlook}\label{sec:5} 
In the present paper, we have demonstrated that despite the absence of NP in
$\varepsilon_K$ large NP effects can be found in
$\kpn$, $\klpn$, $K_S\to\mu^+\mu^-$, $K_L\to\pi^0\ell^+\ell^-$, $\epe$ and $\Delta M_K$. For this to happen the flavour changing coupling must be close to the imaginary one, reducing the number of free parameters. As the CKM parameters have
been determined precisely from $\Delta F=2$ observables only \cite{Buras:2022wpw}, the paucity of NP parameters in this scenario implies strong correlations between all observables involved.

In the coming years, the most interesting will be an improved measurement of the $\kpn$ branching ratio, 
which if different from the SM prediction will imply NP effects in the remaining
observables considered by us. Figs. \ref{Fig:2} and \ref{Fig:3}.
illustrate this in a spectacular manner.  The size of possible enhancements will depend on the involved couplings, in particular, leptonic ones so that correlations with $B$ physics observables
will also be required to get the full insight into the possible anomalies. With improved theory estimates of $\epe$ and $\Delta M_K$, the improved measurements of $\klpn$ and $K_S\to\mu^+\mu^-$, and those of $B$ decays, this
simple scenario will undergo very strong tests.

\medskip
  
 {\bf Acknowledgements}
J.A. has received funding from
 the European Research Council (ERC) under the European Union’s Horizon 2020 research and
 innovation programme under grant agreement 833280 (FLAY), and by the Swiss National Science
 Foundation (SNF) under contract 200020 204428.
 Financial support for A.J.B from the Excellence Cluster ORIGINS,
funded by the Deutsche Forschungsgemeinschaft (DFG, German Research
Foundation), Excellence Strategy, EXC-2094, 390783311 is acknowledged. 
{Research (J.K.) presented in this article was supported by the Laboratory Directed Research and 
Development program of Los Alamos National Laboratory under project number 20220706PRD1.

\addcontentsline{toc}{section}{References}

\small

\bibliographystyle{JHEP}
\bibliography{Bookallrefs}

\end{document}